\documentclass[11pt]{article}

\usepackage{array,makecell}

\usepackage{amsfonts,amssymb,amsthm,bbm,mathtools}
\usepackage{fullpage}
\usepackage{caption}
\usepackage[colorlinks=true,linkcolor=blue,citecolor=magenta,linktocpage=true]{hyperref}
\usepackage{braket}
\usepackage{titlesec}
\usepackage[usenames,dvipsnames]{xcolor}

\usepackage[mathscr]{euscript}
\usepackage{dsfont}
\usepackage{slashed}
\usepackage{currvita}
\usepackage{tocloft}

\usepackage{amsmath}
\usepackage{apptools}

\usepackage{subcaption}
\usepackage{color}
\usepackage{graphicx}
\usepackage{tikz}
\usetikzlibrary{calc}
\usetikzlibrary{decorations.pathmorphing}
\usetikzlibrary{decorations.markings}
\usepackage[numbers,sort,compress]{natbib}

\makeatletter
\renewcommand{\@dotsep}{1000}

\titleformat{\section}{\center\large\bfseries}{\IfAppendix{\appendixname}{} \thesection}{1em}{}

\titleformat*{\subsection}{\flushleft\bfseries}

\definecolor{airforceblue}{rgb}{0.36, 0.54, 0.66}
\definecolor{antiquefuchsia}{rgb}{0.57, 0.36, 0.51}
\definecolor{blush}{rgb}{0.87, 0.36, 0.51}
\definecolor{bondiblue}{rgb}{0.0, 0.58, 0.71}
\definecolor{MyGreen}{rgb}{0.0,0.5,0}
\definecolor{MyDarkRed}{rgb}{0.7,0,0}
\definecolor{MyBlue}{rgb}{0.0,0.0,.5}
\def\be#1\ee{\begin{align}#1\end{align}}
\def\bsub#1\esub{\begin{subequations}#1\end{subequations}}
\def\bg#1\eg{\begin{gather}#1\end{gather}}
\def\ba{\begin{eqnarray}}
\def\ea{\end{eqnarray}}

\def\q{\qquad}
\def\f{\frac}

\def\lb{\big\lbrace}
\def\rb{\big\rbrace}

\def\ip{\lrcorner\,}
\def\ipp{\ip\!\!\!\ip}

\def\rm#1{\mathrm{#1}}

\def\de{\mathrm{d}}

\newcommand{\R}{{\mathbb R}}

\newcommand{\cD}{{\mathcal D}}
\newcommand{\cE}{{\mathcal E}}
\newcommand{\cF}{{\mathcal F}}

\newcommand{\cL}{{\mathcal L}}
\newcommand{\cM}{{\mathcal M}}

\newcommand{\cO}{{\mathcal O}}

\newcommand{\cQ}{{\mathcal Q}}

\newcommand{\cS}{{\mathcal S}}
\newcommand{\cT}{{\mathcal T}}

\newcommand{\cV}{{\mathcal V}}

\newcommand{\bms}{{\mathfrak{bms}}}

\numberwithin{equation}{section}

\begin{document}

\title{\Large{\textbf{\sffamily BMS$_{\boldsymbol{3}}$ Mechanics and the Black Hole Interior}}}
\author{\sffamily Marc Geiller, Etera R. Livine, Francesco Sartini}
\date{\small{\textit{Univ Lyon, ENS de Lyon, Univ Claude Bernard Lyon 1,\\ CNRS, Laboratoire de Physique, UMR 5672, F-69342 Lyon, France\\}}}

\maketitle

\begin{abstract}
The spacetime in the interior of a black hole can be described by an homogeneous line element, for which the Einstein--Hilbert action reduces to a one-dimensional mechanical model. We have shown in \cite{Geiller:2020xze} that this model exhibits a symmetry under the $(2+1)$-dimensional Poincar\'e group. Here we explain how this can be understood as a broken infinite-dimensional BMS$_3$ symmetry. This is done by reinterpreting the action for the model as a geometric action for BMS$_3$, where the configuration space variables are elements of the algebra $\mathfrak{bms}_3$ and the equations of motion transform as coadjoint vectors. The Poincar\'e subgroup then arises as the stabilizer of the vacuum orbit. This symmetry breaking is analogous to what happens with the Schwarzian action in AdS$_2$ JT gravity, although in the present case there is no direct interpretation in terms of boundary symmetries. This observation, together with the fact that other lower-dimensional gravitational models (such as the BTZ black hole) possess the same broken BMS$_3$ symmetries, provides yet another illustration of the ubiquitous role played by this group.
\end{abstract}

\thispagestyle{empty}
\newpage
\setcounter{page}{1}

\hrule
\vspace{-0.3cm}
\tableofcontents
\addtocontents{toc}{\protect\setcounter{tocdepth}{3}} 
\vspace{0.5cm}
\hrule

\newpage

\section{Introduction}

Symmetries are an invaluable tool when studying classical theories and their quantization. They help understand the structure of the classical field equations and of their solutions, and representation theory of symmetry groups organize the states and spectra at the quantum level. It is clear that the quantization of gravity, if attainable, will require a profound understanding of the role of symmetries in general relativity.

General relativity being a gauge theory, the role of symmetries therein is particularly subtle. It has been known for a while that gauge symmetries acquire a particular status in the presence of boundaries \cite{Regge:1974zd,Lee:1990nz,Iyer:1994ys,Wald:1999wa,Balachandran:1994vi,Balachandran:1995qa,Balachandran:1994up}. Boundaries indeed have the ability of turning gauge symmetries into physical symmetries with a non-trivial charge. In the gravitational context, this is related to the appearance of infinite-dimensional boundary symmetry groups, whose first historical examples are the BMS$_4$ enhancement of Poincar\'e in 4-dimensional asymptotically flat spacetimes \cite{Bondi:1962px,Sachs:1962zza,Sachs:1962wk}, and the double Virasoro enhancement of SO(2,2) in AdS$_3$ gravity \cite{Brown:1986nw}. In certain cases these boundary symmetries can moreover be seen as the bulk symmetries of a lower-dimensional theory describing the boundary dynamics \cite{Coussaert:1995zp,Barnich:2015sca,Barnich:2013yka,Barnich:2012rz}. 

The particular case of the BMS$_4$ group has received lots of attention in recent years due to newly discovered connections with gravitational scattering and memory effects \cite{Strominger:2013jfa,He:2014laa,Compere:2016gwf,Choi:2017ylo,Ashtekar:2018lor,Rahman:2019bmk}, and the prospect of constructing holographic theories \cite{Arcioni:2003xx,Arcioni:2003td,Barnich:2010eb,Fotopoulos:2019vac,Laddha:2020kvp,Donnay:2020guq,Puhm:2019zbl,Guevara:2021abz}. This has motivated in particular authors to propose various extensions of the 4-dimensional BMS group \cite{Barnich:2010eb,Barnich:2011mi,Flanagan:2015pxa,Ruzziconi:2020cjt,Compere:2020lrt,Freidel:2021yqe,Campiglia:2021bap}. The 3-dimensional group BMS$_3$ and its extensions have also been studied at length \cite{Ashtekar:1996cd,Barnich:2006av,Barnich:2012aw,Oblak:2016eij,Carlip:2017xne,Barnich:2014kra,Barnich:2015uva,Barnich:2017jgw,Penna:2017vms,Carlip:2019dbu,Ruzziconi:2020wrb,Geiller:2021vpg}. In all these works the BMS group is always associated with the symmetries of some boundary structure, whether at infinity or on a black hole horizon. In the present paper we study a setup where the group BMS$_3$ plays a role, although as far as we can tell there are no boundaries or asymptotic symmetries involved.

To set the stage, let us recall that in a recent paper \cite{Geiller:2020xze} we have studied the dynamics of the black hole interior spacetime. This is described by an homogeneous line element, the Kantowski--Sachs metric, for which the Einstein--Hilbert action reduces to a 1-dimensional mechanical model, with time reparametrization symmetry inherited from the 4-dimensional diffeomorphism invariance. We have shown that the resulting action (and the Hamiltonian theory), when gauge-fixed to proper time, exhibits a residual invariance under the $(2+1)$-dimensional Poincar\'e group ISO$(2,1)$. This extends the results obtained for FLRW spacetimes in \cite{BenAchour:2017qpb,BenAchour:2019ywl,BenAchour:2019ufa,BenAchour:2020njq,BenAchour:2020xif}.

The aim of the present work is to study how this Poincar\'e symmetry group in the black hole interior can be embedded into its infinite-dimensional enhancement BMS$_3$, and how the results of \cite{Geiller:2020xze} can be reinterpreted from this point of view. As we will see, even if the theory is not invariant under the full group but only under its global Poincar\'e subgroup, it is possible to associate integrable generators to all the BMS$_3$ transformations on the phase space. We then show that the classical trajectories can be interpreted in terms of coadjoint orbits of BMS$_3$. More precisely, the equations of motion themselves transform as coadjoint vectors. Motivated by this observation, we show that an equivalent action for the black hole interior can be obtained in the form of a geometric action for BMS$_3$, i.e. as a pairing between coadjoint vectors (the equations of motion) and elements of the Lie algebra $\mathfrak{bms}_3$ (the configuration variables). With this interpretation the theory picks up the gravitational central charge $c_2$ of BMS$_3$. The new action resembles closely the charges of the 2-dimensional BMS--Liouville field theory introduced in \cite{Barnich:2012rz,Barnich:2013yka}, although as we have mentioned the 1-dimensional black hole interior action is not BMS$_3$ invariant and is not obtained as a boundary theory. Instead, there is a symmetry breaking and only the Poincar\'e subgroup survives as the symmetry group of the theory.


It is known however that there are setups in which the boundary, instead of enlarging the symmetries (to e.g. BMS), partially breaks the symmetry to a subgroup. This is for example the case of the Schwarzian theory living at the AdS boundary of JT gravity \cite{Almheiri:2014cka,Maldacena:2016upp,Engelsoy:2016xyb,Jensen:2016pah,Godet:2021cdl}, where the original Virasoro group is broken to its global SL$(2,\R)$ part. All these observations suggest that we are actually dealing with a symmetry breaking inherited from a higher-dimensional theory, potentially related to the boundary geometric actions appearing in 3-dimensional gravity, but the clear relationship is still missing. What is also surprising is that although this residual Poincar\'e invariance (and the related BMS interpretation) has been initially found in the 4-dimensional black hole interior, it turns out that there exist a broad class of other minisuperspace models in which the same 1-dimensional action appears. This is for example the null slicing of the Schwarzschild solution, the BTZ black hole interior, and a class of 2-dimensional dilatonic gravity theories. This strongly suggests that there is a universal yet still elusive mechanism at play and involving the aforementioned symmetry groups.

The paper is organised as follows. In section \ref{sec:bh_interior} we review the classical setup for the study of the homogeneous black hole interior spacetime. We explain in particular the choice of lapse, the gauge fixing to proper time, and show how the solutions to the equations of motion reconstruct the black hole interior geometry. Then we move on in the next sections to the study of the symmetries of the model. In \ref{sec:bms_adjoint} we define the adjoint representation of BMS, and use this to act with finite and infinitesimal transformations on the action and to derive the integrable Hamiltonian generators. This shows that the Poincar\'e subgroup indeed defines symmetries, and also shows how a fully BMS$_3$ invariant theory can be obtained via a Stueckelberg mechanism. This fully invariant theory is further studied in appendix \ref{Full_invariant}. In section \ref{coadj_section} we then turn to the coadjoint representation of BMS$_3$, and explain how the equations of motion and the broken Poincar\'e symmetry can be understood in terms of coadjoint orbits. The main result is the new form \eqref{BMS_geometric_action} of the action, which is a pairing between coadjoint vectors and the configuration space variables which transform as elements of the Lie algebra $\mathfrak{bms}_3$. Appendix \ref{BMS_appendix} contains various definitions about the centrally-extended Virasoro and BMS$_3$ groups. In appendix \ref{other_slicing} we present other systems in 4, 3, and 2-dimensional gravity which are governed by the same action as the 4-dimensional black hole interior, and therefore admit the same symmetry structure.

\section{The black hole interior spacetime}
\label{sec:bh_interior}

In this first section we present the classical model underlying our construction, namely the homogeneous black hole interior spacetime. We start by explaining the choice of lapse and clock, and analyse the resulting Hamiltonian. We then present the symmetry reduced action in proper time and recall how the resulting equations of motion reconstruct the black hole interior spacetime.

\subsection{Homogeneous metric, choice of time and Hamiltonian}

The region inside the horizon of a Schwarzschild black hole of mass $M$ is described by the  Kantowski--Sachs metric, with the homogeneous though anisotropic line element
\be\label{Schwarz_interior}
\de s^2 = -\left (\f{2M}{T}-1\right )^{-1} \de T^2 + \left (\f{2M}{T}-1\right )\de r^2 + T^2 \de \Omega^2\,,
\ee 
where $\de \Omega^2$ is the metric on the unit 2-sphere at constant $r$ and $T$. In these coordinates, the singularity is located at $T=0$ and the horizon is at $T=2M$. In order to study the dynamics of this system, its symmetries, perturbations around this classical solution and their possible quantization, we introduce dynamical fields and consider the more general homogeneous ansatz for the line element:
\be
\label{metric_vp}
\de s^2 = -N(t)^2 \de t^2 + \f{8 V_2(t)}{V_1(t)} \de x^2 + V_1(t) \de \Omega^2\,,
\ee
where the spatial coordinate runs over the whole real line, $x\in \R$. The spatial slices have the topology $\R \times S^2$. Because of the spatial slices are non-compact, the evaluation of the Einstein--Hilbert action on the line element \eqref{metric_vp} diverges, due to the constant integration over $x$. This can be regularized with the introduction of a fiducial length $L_0$ playing the role of an IR cutoff. Restricting the integration range to $x\in [0,L_0]$, the vacuum Einstein--Hilbert action evaluated on \eqref{metric_vp} becomes
\be
\label{Einstein--Hilbert}
\cS_\text{EH}^{(t)}[N,V_i]
&=\f{1}{16\pi G}\int_{\cM} \de^4 x\, \sqrt{-g}\,R\cr
&=\f{1}{G}\int_{\R} \de t\,L_0 \left[\f{V_2 (4 N^2 V_1 + V_1'^2) -  2 V_1  V_1'  V_2'}{2 \sqrt 2 N (V_1^3  V_2)^{1/2}}+\f{\de}{\de t} \left (\f{1}{\sqrt{2}N} \big(\sqrt{V_1  V_2}\big)' \right )\right]\,,
\ee
where the prime denotes the derivative with respect to coordinate time $t$. The total derivative is the Gibbons--Hawking--York term associated with slices at constant $t$. We will therefore drop this term since it does not play a role in the study of the equations of motion and does not affect the discussion of the symmetries.

In a previous work \cite{Geiller:2020xze} we have shown that the analysis of the classical symmetries of this 1-dimensional mechanical action is more transparent when redefining the lapse as
\be\label{lapse Np}
N =L_0N_\text{p} \sqrt{\f{ V_1}{2  V_2}}\,, \q\q
\de s^2=-L_0^2N_\text{p}(t)^2\f{V_1(t)}{2V_2(t)}\de t^2+\f{8V_2(t)}{V_1(t)}\de x^2+V_1(t)\de\Omega^2\,.
\ee
We emphasize that this redefinition actually corresponds to a field-dependent and $L_0$-dependent time diffeomorphism\footnotemark{}.
\footnotetext{This is possible because we do not impose any boundary condition on the lapse, but it nevertheless begs the question of the equivalence between the two gauges, especially at quantum level.} The action  now reads
\be
\label{EHt}
\cS_\text{EH}^{(t)}[N_\text{p},V_i]
&=
\f{1}{G}\int \de t \, \left[N_\text{p}L_0^2 + \f{ V'_1 ( V_2 V'_1 - 2  V_1  V'_2)}{2N_\text{p}V_1^2 }\right]
\,.
\ee
Then, we introduce the \textit{proper time} gauge by working with $\de\tau = N_\text{p} \de t$. With this redefinition of the lapse and the choice of proper time, the lapse disappears from the action, which  becomes
\be
\cS_\text{EH}^{(\tau)}[N_\text{p},V_i]
=\f{1}{G}\int \de \tau \, \left[L_0^2 + \f{ \dot{V}_1 ( V_2 \dot{V}_1 - 2  V_1  \dot{V}_2)}{2V_1^2 }\right]\,,
\ee
where the dot denotes the derivative with respect to proper time $\tau$. The contribution of $L_0$ is now a boundary term. As such, it cannot contribute to the equations of motion and should simply be discarded from the action. We therefore redefine the action in proper time as
\be\label{new_lapse_action}
\cS_0[V_i]
\equiv
\f{1}{G}\int \de \tau \,\cL_0=\f{1}{G}\int \de \tau \,\f{ \dot{V}_1 ( V_2 \dot{V}_1 - 2  V_1  \dot{V}_2)}{2V_1^2 }\,.
\ee
On the one hand, one might think that this is a great feature of the choice of time $\tau$. Indeed, this gauge fixing of the time reparametrization symmetry to proper time not only removes the lapse but also gets rid of the IR cut-off $L_0$. On the other hand, this IR scale is simply a pre-factor in front of the whole action $\cS_\text{EH}^{(t)}[N,V_i]$, so how can it disappear?

This intriguing property is deeply intertwined with the role of the lapse. Indeed, even if the lapse has formally disappeared from the action $\cS_\text{EH}^{(\tau)}$, it still enters implicitly in the definition of $\tau$. In fact, the equation of motion resulting from the stationarity of the Lagrangian in $\cS_\text{EH}^{(t)}$ with respect to variations of the lapse translates into the fact that the Hamiltonian of $\cS_\text{EH}^{(\tau)}$ dictating the evolution in proper time $\tau$ is a constant of motion and is equal to $L_0^2$. Thus the Hamiltonian of the action in proper time does not vanish and the term $L_0^2$ is now interpreted as a non-zero level for this Hamiltonian which generates the evolution with respect to $\tau$ \cite{Geiller:2020xze}. 

Let us check this in more details. Differentiating the action $\cS_\text{EH}^{(t)}[N_\text{p},V_i]$ with respect to the redefined lapse $N_\text{p}$ gives the following equation of motion:
\be
L_0^2=\f1{N_\text{p}^2}\left[\f{V_2{V'_1}^2}{2V_1^2}-\f{V'_1V'_2}{V_1}\right]
=\f{V_2\dot{V}_1^2}{2V_1^2}-\f{\dot{V}_1\dot{V}_2}{V_1}\,.
\ee
Now, performing the canonical analysis of the Lagrangian $\cL_0$ gives the following conjugate momenta:\footnote{Note that for convenience we perform the Legendre transform of $\cL_0$ alone and leave out the prefactor $1/G$.}
\be
P_1=\f{\partial \cL_0}{\partial \dot{V}_1}=\f{V_2\dot{V}_1-V_1\dot{V}_2}{V_1^2}\,,\q\q
P_2=\f{\partial \cL_0}{\partial \dot{V}_2}=-\f{\dot{V}_1}{V_1}\,,
\ee
and the following Hamiltonian:
\be
H=P_1\dot{V}_1+P_2\dot{V}_2-\cL_0
=
\f{V_2\dot{V}_1^2}{2V_1^2}-\f{\dot{V}_1\dot{V}_2}{V_1}
= -P_2 \left(V_1P_1+\f{1}{2} V_2 P_2 \right)\,.
\ee
It is then obvious that the equation of motion for the lapse $N_\text{p}\propto N$ is exactly equivalent to fixing the value of the energy for the evolution in proper time $\tau$, i.e.
\be
\f{\delta \cS_\text{EH}^{(t)}}{\delta N_\text{p}}=0\quad\Leftrightarrow\quad
H=L_0^2\,.
\ee
This can at first seem like a surprising role for the IR cut-off as the Hamiltonian level for the evolution in proper time, but we have shown that this role is actually very natural.

Thus, at the end of the day, we can work in proper time and forget about the lapse $N$ and the IR cut-off $L_0$, as long as we remember to fix the value of the Hamiltonian to $L_0^2$ when inserting the on-shell fields in the line element.


 
\subsection{Reduced action and classical solutions}

The starting point of the present work is the 1-dimensional action \eqref{new_lapse_action} written in proper time $\tau$. The variation of this action can be written in the compact form
\be
\label{variation_action_0}
\delta \cS_0=\f{1}{G}\int \de \tau\,\big[ J\delta V_1+P\delta V_2+\de_\tau\theta\big]\,,
\ee 
with
\bsub\label{J P theta}
\be
J&\coloneqq\f{V_2 \dot{V}_1^2}{V_1^3} -\f{\dot{V}_1 \dot{V}_2}{V_1^2}- \f{V_2 \ddot V_1}{V_1^2} +\f{\ddot V_2}{V_1},\label{J def}\\
P&\coloneqq\f{\ddot V_1 }{V_1}- \f{\dot{V}_1^2}{2 V_1^2},\label{P def}\\
\theta&\coloneqq\f{\dot{V}_1 V_2 - \dot{V}_2 V_1}{V_1^2} \delta  V_1-\f{\dot{V}_1}{V_1} \delta V_2.\label{V potential}
\ee
\esub
The equation of motion are thus given by $J=0$ and $P=0$, and $\theta$ is the pre-symplectic potential. As we will see in section \ref{coadj_section}, the names $J$ and $P$ have suggestively been chosen because of the relationship between the equations of motion and the coadjoint representation of BMS$_3$. One can notice that by removing the boundary term in \eqref{Einstein--Hilbert} (which is equivalent to adding the GHY term), we have eliminated the second order time derivatives from the action and obtained a well-defined variational principle for fixed fields $V_i$ at initial and final time (i.e. a fixed metric).

We can now look at the solutions of the equations of motion $J=0=P$. They are given by
\bsub\label{classic solution}
\be
V_1 &= \f{A}{2} (\tau-\tau_0)^2\,,\\
V_2 &= B(\tau-\tau_0) -\f{L_0^2}{2} (\tau-\tau_0)^2\,,\ee
\esub
where $A$, $B$ and $\tau_0$ are constants of integration.
In these solutions the constant $\tau_0$ reflects the freedom in choosing the origin for the time variable, and can be set to zero without loss of generality. The constants $A$ and $B$ are integrals of motion, which can be identified as:
\be
A=\f{\dot{V}_1^2}{2V_1}
\,,\qquad
B=\f{V_2\dot{V}_1-V_1\dot{V}_2}{V_1}
\,.
\ee
Finally, $L_0$ appears because we have used the constraint initially enforced by the lapse $N_\text{p}$ and fixing the energy level of the Hamiltonian, i.e.
\be
\label{energy_L0}
\f{\dot{V}_1 (V_2 \dot{V}_1- 2 V_1 \dot{V}_2)}{2V_1^2} =L_0^2\,.
\ee
Inserting these solution into the line element \eqref{lapse Np}, and changing the variables as 
\be
\tau -\tau_0 = \sqrt{\f{2}{A}} T\,,\q\q x=\f{1}{2 L_0}\sqrt{\f{A}{2}} r\,,
\ee
we recover the Schwarzschild interior solution \eqref{Schwarz_interior} with mass
\be 
M=\f{B \sqrt{A}}{\sqrt{2} L_0^2}\,.
\ee
This provides the physical interpretation for the constants of motion $A$ and $B$.

Before moving on to the study of the BMS structure, we would like to point out that the symmetry-reduced action $\cS_0$ is not specific to the case of the 4-dimensional black hole interior, but emerges also in various other mini-superspace gravitational systems. Interestingly, some of these systems are not 4-dimensional. For instance, we find $\cS_0$ for a null slicing of the 4d black hole spacetime, for an homogeneous slice of the 3d BTZ black hole interior, and for a large class of 2d dilatonic gravity models. These examples are briefly presented in appendix \ref{other_slicing}. This strongly hints at the presence of some universal and dimension-independent structure, although its precise characterization remains elusive as for now.

Moreover, despite the fact that $\cS_0$ has been found in relationship with the black hole interior, and should be in principle limited to the range in which $\tau$ is timelike, the fields $V_i$ are regular on the whole real line. The \textit{classical} evolution smoothly cross not only the horizon, but also the singularity. 

\section{Adjoint representation of BMS${}_3$}
\label{sec:bms_adjoint}

We now set out to explain the relationship between the action \eqref{new_lapse_action} and BMS${}_3$. In our previous work \cite{Geiller:2020xze} we have shown that the dynamical system describing the black hole interior via \eqref{new_lapse_action} admits a set of symmetries whose finite form is isomorphic to the 3-dimensional Poincar\'e group ISO$(2,1)$. Moreover, we have shown that these symmetries can be realized by the embedding of the $2+1$ Poincar\'e group into the larger\footnote{Later on we will use the centrally-extended group BMS$_3$. The central extensions are defined in appendix \ref{BMS_appendix}, where we use the notation $\widehat{G}$ for the centrally-extended group $G$. In the core of the text however we drop this hat in order to deal with lighter notations.} $\text{BMS}_3=\text{Diff}(S^1) \ltimes \text{Vect}(S^1)_\text{ab}$  group, where $\text{Diff}(S^1)$ is the group of diffeomorphism of the (compactified) time axis, while $\text{Vect}(S^1)_\text{ab}$ is its Lie algebra seen as an Abelian vector group. The natural question is now to understand the reformulation of the Kantowski--Sachs mini-superspace model in terms of the (centrally-extended) BMS group, and how the BMS symmetry is broken down to Poincar\'e. For this we will first discuss in details the symmetries of the action \eqref{new_lapse_action} using the adjoint representation, and then explain in the next section the relationship with coadjoint representations.
We refer the reader to appendix \ref{BMS_appendix} for more details on the properties of the (centrally-extended) BMS$_3$ group and its representations.

\subsection{Symmetries and transformations of the action}

The previous work \cite{Geiller:2020xze} started from the Poincar\'e symmetry of the Kantowski--Sachs reduced action and, trying to extend it to a BMS symmetry, realized that the metric components $V_1$ and $V_2$ transform as elements of the Lie algebra $\mathfrak{bms}_3$ in the adjoint representation of the BMS group. Let us recall how this comes about.

We consider transformations $D_f\in\text{Diff}(S^1)$ and $T_g\in\text{Vect}(S^1)_\text{ab}$, and define their action on the configuration variables $V_i=(V_1,V_2)$ as
\be\label{fgreparam}
D_{f}:\,&\left|
\begin{array}{lcl}
V_i &\mapsto& \widetilde V_i = (\dot{f}\, V_i)\circ f^{-1}\,,
\end{array}
\right.\\
T_{g}:\,&\left|\begin{array}{lcl}
V_{1}&\mapsto& \widetilde{V}_{1}=V_{1}\,,\\
V_2 &\mapsto& \widetilde {V}_2  = V_2 + g\dot{V}_1  -  \dot{g}V_1\,,
\end{array}\right.
\ee
where we recall that all these quantities depend on $\tau$. These are the transformation laws derived in \cite{Geiller:2020xze}. They are given here in a form acting only on the dynamical fields $V_i$. To compute the transformation of the action it will be convenient to act instead with $D_f$ equivalently rewritten as
\be\label{freparam}
D_{f}:\,\,\left|
\begin{array}{lcl}
\tau &\mapsto& \tilde{\tau} = f(\tau)\,,\vspace*{1mm}\\
V_i &\mapsto& \widetilde V_i (\tilde{\tau}) = \dot f(\tau)\, V_i(\tau)\,,
\end{array}
\right.
\qquad\textrm{so}\q
\left|
\begin{array}{lcl}
\de \tau &\mapsto& \de \tilde{\tau} = \dot f \de \tau
\,,\vspace*{1mm}\\
 {\de_{\tau} V_i } &\mapsto&
\de_{\tilde{\tau}} \widetilde V_i = {\de_{\tau} V_i } + V_i\,\de_{\tau}\ln \dot f
 \,.
\end{array}
\right.
\ee
Following the terminology introduced for asymptotic symmetries, we will (abusively) refer to the transformations $D_f$ as \textit{superrotations} and to $T_g$ as \textit{supertranslations}.
In order to justify this name, we now show that the composition
\be\label{Ad definition}
\text{Ad}_{f,g}\coloneqq T_{g}\circ D_{f}
\ee
does indeed define a representation of BMS$_3$. For this, we use the properties
\be
D_{f^{-1}}\circ T_{g}\circ D_{f}=T_{(g\circ f)/\dot{f}},
\q\q
D_{f_1}\circ D_{f_2}=D_{f_1\circ f_2}\,, \q\q
T_{g_1}\circ T_{g_2}=T_{g_1+g_2}\,,
\ee
which allow us to show that the composition law for \eqref{Ad definition} is given by
\be\label{BMScomposition}
\text{Ad}_{f_1,g_1}\circ \text{Ad}_{f_2,g_2}
&= T_{g_1}\circ D_{f_1}\circ T_{g_2}\circ D_{f_2}\cr
&=T_{g_1}\circ (D_{f_1}\circ T_{g_2}\circ D_{f_1}^{-1}) \circ (D_{f_1}\circ D_{f_2}) \cr
&=T_{g_1+\dot{f}_1\,g_2 \circ f_1^{-1}}\circ D_{f_1\circ f_2}. 
\ee
The inverse composition is
\be
(T_{g}\circ D_{f})^{{-1}}
=
D_{f^{-1}}\circ T_{-g}
=
T_{-(g\circ f)/\dot{f}}\circ D_{f^{-1}}
\,.
\ee
As desired, this is the group multiplication and its inverse for the Lie group defined as the semi-direct product $\text{Diff}(S^1) \ltimes \text{Vect}(S^1)_\text{ab}$.

The interpretation is then that the gravitational variables $V_1$ and $V_2$ belong to the Lie algebra
\be
\mathfrak{bms}_3=\text{Vect}(S^1) \oplus_\text{ad} \text{Vect}(S^1)_\text{ab}\ni(V_1,V_2)\,,
\ee
and are acted on by the group via the adjoint action \eqref{BMS_adjoint}.
 
\subsubsection{Finite transformations}

We now study how the action \eqref{new_lapse_action} transforms under finite BMS transformations. It turns out to be easier to compute this using the inverse transformations to \eqref{fgreparam}. Acting jointly with superrotations and supertranslations we find
\be
\label{transfo_action}
\text{Ad}_{(f^{-1},g)} \cS_0[V_i,\tau] &=\cS_0[V_i,\tilde \tau] - \f{1}{G}\int \de \tilde \tau\left [\vphantom{\f{1}{G}}(V_2-\dot{g}V_1+g \dot{V}_1)\,\text{Sch}[f^{-1}]-V_1g^{(3)} \right.\\
&\phantom{=\cS_0[V_i, \tilde \tau] +\f{1}{G} \int \de \tilde \tau\left[\right.~}+\left.\f{\de}{\de\tilde{\tau}} \left (\f{\ddot f}{\dot{f}}(V_2-\dot{g}V_1+g \dot{V}_1)+ \ddot{g}V_1 - \f{g \dot{V}_1^2}{2V_1} \right ) \right ]\,,\notag
\ee
where $\text{Sch}[\,\cdot\,]$ denotes the Schwarzian, whose properties are recalled in appendix \ref{BMS_appendix}. As expected, this expression has cross terms involving both $f$ and $g$ since we have acted jointly with $\text{Ad}_{f,g}$. Focusing on the sector $f$, it is immediate to see on this expression that the subgroup of transformations with vanishing Schwarzian derivative are symmetries of the reduced gravitational action. These are given by M\"obius transformations, isomorphic to the group SL$(2,\R)$. Similarly, in the $g$ sector we obtain a symmetry if $g$ has a vanishing third derivative, meaning that it is a second degree polynomial. This shows that the subgroup of \eqref{fgreparam} which describes the symmetries of \eqref{new_lapse_action} is given by $\text{SL}(2,\R)\ltimes\R^3$, which is the $2+1$ Poincar\'e group \cite{Geiller:2020xze}.

When the transformation $\text{Ad}_{f,g}$ does not belong to the Poincar\'e subgroup, we see that the general BMS transformation produces terms proportional to the dynamical variables $V_1$ and $V_2$ in the action. As explained in appendix \ref{lambda_and_phi}, terms in $V_1$ and $V_2$ also appear when introducing a cosmological constant or a scalar field in the model. While it is tempting to interpret the BMS transformation as producing a cosmological constant term or a scalar field, this interpretation does not strictly hold since, as can be seen on \eqref{scalar field in S} a scalar field comes with specific time derivatives which cannot be obtained from $f$ and $g$ in \eqref{transfo_action}. However, in the case of pure supertranslations $g$ (i.e. with $f=\mathbb{I})$ one can really interpret the term $g^{(3)}V_1$ as the contribution of a cosmological constant. The action of a further finite supertranslation then shifts the value of this cosmological constant.

As an alternative to this interpretation in terms of a cosmological constant and a scalar field, we can also create a fully BMS invariant action by including extra fields in the theory. These have to transform in a way which compensates the Schwarzian and the $g^{(3)}$ produces by a finite BMS transformation. This is achieved by the invariant action
\be
\label{invariant_1d_action}
\cS_\text{inv}[V_i,\Phi,\Psi] \coloneqq\cS_0[V_i] + \f{1}{G}\int  \de \tau\left [(V_2+ \dot \Psi V_1- \Psi \dot{V}_1) \text{Sch}[\Phi]  + V_1\Psi^{(3)}  \right ],
\ee
provided the new fields $\Phi$ and $\Psi$ (which can be seen as St\"uckelberg fields for the BMS symmetry) transform as
\be
\label{FGreparam}
\text{Ad}_{f,g}:\,&\left|
\begin{array}{lcl}
\Phi &\mapsto& \widetilde \Phi = \Phi \circ f^{-1}\,,\\
\Psi &\mapsto& \widetilde \Psi = g + (\dot{f} \Psi)\circ f^{-1}\,.\\
\end{array}
\right.
\ee
This BMS invariant action can be seen as the extension from the Virasoro group to the BMS group of the conformally-invariant action for FLRW cosmology that was introduced in \cite{BenAchour:2020xif}. The cosmological model can be understood as the $V_1=0$ regime of the theory with $V_2$ playing the role of the cosmological volume. In that case, there are no supertranslations but only superrotations and the conformally-invariant cosmological action was defined by introducing solely an extra $V_2\,\text{Sch}[\Phi]$ term.

In terms of group representation, this means that the fields are embedded in the BMS${}_3$ group itself:
\be
(\Phi,\Psi)\in\text{BMS}_3=\text{Diff}(S^1) \ltimes_\text{Ad} \text{Vect}(S^1)_\text{ab}.
\ee
This will inevitably modify the equations of motion for the initial fields $V_i$, and we postpone the analysis of this fully invariant theory to appendix \ref{Full_invariant}.

Note that here we have simply introduced the BMS invariant action as an example of a 1-dimensional system with full BMS symmetry. This is to be contrasted with the 2-dimensional BMS--Liouville theory \eqref{BMS Liouville} which will also have a close relationship with the action $\cS_0$ we started from.

\subsubsection{Infinitesimal transformations}

We now move on to the study of the infinitesimal transformations. Since the supertranslations are Abelian, their infinitesimal generator, which we will denote $\alpha$, acts in the same way as the finite generator $g$. For the superrotations, we parametrize the transformation infinitesimally as $f(\tau)= \tau + \epsilon X(\tau)$. Then the infinitesimal version of \eqref{fgreparam} reproduces the $\mathfrak{bms}_3$ algebra \eqref{BMS_algebra}. We then obtain
\be\label{X_infinit}
\delta_X V_i\coloneqq\frac{\text{Ad}_{f,0} V_i -V_i}{\epsilon}= \dot XV_i-  X\dot{V}_i = [V_i,X],
\ee
and
\bsub\label{a_infinit}
\be
\delta_\alpha V_1 &= 0,\\
\delta_\alpha V_2 &= \alpha \dot{V}_1- \dot \alpha V_1 = -[V_1,\alpha].
\ee
\esub
With this, the infinitesimal variation of the action is
\be
\delta_{X,\alpha} \cS_0 = \f{1}{G}\int \de \tau \left[V_2 X^{(3)}-V_1 \alpha^{(3)} \right ] + \de\tau\f{\de}{\de\tau}\left[\f{\dot{V}_1}{2V_1^2}\Big(2XV_1\dot{V}_2-\dot{V}_1(\alpha V_1+XV_2)\Big)+\ddot{\alpha}V_1-\ddot{X}V_2\right].
\ee
Once again, we see that the Poincar\'e subgroup, which at the infinitesimal level has generators such that $X^{(3)}=0=\alpha^{(3)}$, generates symmetries of the system since this variation then reduces to a total derivative.

\subsection{Hamiltonian formulation and generators}
\label{subsec_hamilt_gen}

Having studied the action of finite and infinitesimal BMS transformations on our system $\cS_0$, we can now study the generators in the (covariant) Hamiltonian formulation. From the general variational expression we can read the symplectic potential $\theta$. Its expression can be simplified if we change variables and introduce
\be
\varphi=-\log V_1\,, \q\q \xi = \f{V_2}{V_1}\,.
\ee
In terms of these variables $J$ and $P$ in \eqref{J P theta} become
\be\label{new J P}
P=\f{1}{2}\left (\dot \varphi^2 - 2\ddot \varphi\right )\,,\q\q
J&=\ddot{\xi}-\dot \xi \dot \varphi\,,
\ee
and the equations of motion are $J=0=P$. One can note that in terms of $W$ defined by $\dot{W}=e^\varphi$ we have $P=-\text{Sch}[W]$. The transformations \eqref{fgreparam} become
\be
\label{phixireparam}
\text{Ad}_{f,g}^{-1}:\,&\left|
\begin{array}{lcl}
\varphi &\mapsto& \widetilde \varphi = \varphi \circ f+\log \dot{f}\,,\\
\xi &\mapsto& \widetilde \xi =\left ( \xi + g \dot \varphi + \dot g \right )\circ f\,,\\
\end{array}
\right.
\ee
and we see in particular that under $D_f$ the variables $\varphi$ and $\xi$ transform respectively as a connection and a scalar. The infinitesimal transformations \eqref{X_infinit} and \eqref{a_infinit} take the form
\bsub\label{phixi_infinit}
\be
\delta_{X,\alpha}\, \varphi &= - X \dot \varphi  - \dot X\,,\\
\delta_{X,\alpha}\, \xi &= - X \dot \xi  - \alpha \dot \varphi  - \dot \alpha\,.
\ee
\esub
Rewriting \eqref{V potential} in terms of $\varphi$ and $\xi$, we find the symplectic structure
\be\label{sympl}
\omega =\delta\theta=-\delta \dot \xi\, \delta (e^{-\varphi}) + \delta \dot \varphi\, \delta (\xi e^{-\varphi})\,.
\ee
From this we can rewrite the initial momenta as
\be\label{phase_space}
P_1 = -\dot \xi = \f{\dot{V}_1 V_2 -\dot{V}_2 V_1 }{V_1^2} \,\q \q P_2=\dot \varphi =-\f{\dot{V}_1}{V_1}\,,\q\q\{V_i, P_j\}= \delta_{ij} \,.
\ee
  
With the symplectic structure at our disposal, we can ask whether there are integrable generators associated with the infinitesimal transformations \eqref{phixi_infinit}. This is found by contracting $\omega$ with the variations $\delta_{X,\alpha}$, and then using the equations of motion to get rid of the second order derivatives. Quite surprisingly, even if the general BMS transformations do not describe symmetries of the action, they have an integrable generator. Separating the superrotations and supertranslations, we find
\bsub
\be
\delta_{X} \ipp \omega &= - \delta_{X} \dot \xi\, \delta (e^{-\varphi}) + \delta_{X} \dot \varphi\, \delta (\xi e^{-\varphi}) + \delta \dot \xi\, \delta_{X} (e^{-\varphi}) - \delta \dot \varphi\, \delta_{X} (\xi e^{-\varphi})\cr
&= -\delta \left(e^{-\varphi} \left (\f{\xi}{2}\dot \varphi^2-\dot \varphi \dot \xi \right ) X + e^{-\varphi} (\xi \dot \varphi -\dot{\xi}) \dot X + e^{-\varphi} \xi \ddot X\right )\cr
&= -\delta \left (P_2 \left (P_1 V_1 + \f{1}{2} P_2 V_2\right ) X +(V_2 P_2 +V_1 P_1) \dot X + V_2 \ddot X \right )\cr
&\coloneqq -\delta \cD_X \,,\\
\delta_{\alpha} \ipp \omega &= - \delta_{\alpha} \dot \xi\, \delta (e^{-\varphi}) + \delta_{\alpha} \dot \varphi\, \delta (\xi e^{-\varphi}) + \delta \dot \xi\, \delta_{\alpha} (e^{-\varphi}) - \delta \dot \varphi\, \delta_{\alpha} (\xi e^{-\varphi})\cr
&= \delta \left(\f{1}{2} e^{-\varphi} \dot \varphi^2 \alpha +e^{-\varphi} \dot \varphi \dot \alpha+ e^{-\varphi} \ddot \alpha\right )\cr
&= \delta \left (\f{1}{2} {P_2}^2 V_1 \alpha + P_2 V_1   \dot \alpha + V_1 \ddot \alpha \right )\cr
&\coloneqq -\delta \cT_\alpha \,.
\ee
\esub
It is now interesting to compute the Poisson brackets between these generators. This can be one either by contracting twice the symplectic structure with a variation $\delta_{X,\alpha}$, or using the phase space expression for the generators and the usual definition of Poisson brackets. For example we have
\be
\{\cD_X, \cD_Y\}=- \delta_{X} \ipp \delta_{Y} \ipp \omega = \f{\de \cD_X}{\de V_i}\f{\de \cD_Y}{\de P_i}-\f{\de \cD_X}{\de P_i}\f{\de \cD_Y}{\de V_i}\,. 
\ee
Explicitly, we find
\bsub\label{poisson_bms}
\be
\{\cD_X, \cD_Y\}&=-\cD_{[X,Y]} +\big(X Y^{(3)}-Y X^{(3)} \big) V_2\,,\\
\{\cT_\alpha, \cT_\beta \}&=0\,,\\
\{\cD_X, \cT_\alpha \}&=-\cT_{[X,\alpha]} + \big(\alpha X^{(3)} - X \alpha^{(3)}\big) V_1\,,
\ee
\esub
where $[X,Y]=X\dot{Y}-Y\dot{X}$ and similarly for $[X,\alpha]$.

These brackets are consistent with the equations of motion. For example, one can study the superrotation associated with a constant shift in time. This actually corresponds to the Hamiltonian of the system, i.e. $\cD_{X(\tau)=1}=H$, as one can also check using a Legendre transform of the Lagrangian. Note that this is slightly peculiar since in the usual spacetime picture the Hamiltonian belongs to the Abelian sector of the symmetry algebra (i.e. Poincaré time translations), whereas here it belongs to the superrotations. For a general (time-dependent) phase space function $\cO(\tau)$, the time evolution is given by
\be 
\dot \cO\eqqcolon\de_\tau\cO = \partial_\tau \cO + \{\cO,H\}\,.
\ee
Using the commutation relations \eqref{poisson_bms} applied to $\cD_1$, and the fact that e.g. $\partial_\tau \cD_X = \cD_{\dot X}$, we then find
\bsub
\be
\de_\tau \cD_X &= \partial_\tau \cD_X-\{H,\cD_X\}= V_2 X^{(3)},\\
\de_\tau \cT_\alpha &= \partial_\tau \cT_\alpha-\{H,\cT_\alpha\}= -V_1 \alpha^{(3)},
\ee
\esub
which is consistent with the equations of motion.

In spite of this consistency, we see of course that the bracket \eqref{poisson_bms} between the generators $\cD$ and $\cT$ do not close because of a remaining field-dependency on the right-hand side (although up to the $V_i$'s this would have been the centrally-extended BMS$_3$ algebra). This is to be expected since $\cD$ and $\cT$, although they are integrable, do not generate symmetries of the theory. Consistently, we recover a closed algebra for the Poincar\'e generators since these have a vanishing third derivative.

One way to obtain charges closing on the BMS$_3$ algebra is of course to consider a BMS invariant field theory. Although we have proposed a 1-dimensional such theory in \eqref{invariant_1d_action}, it turns out that a simpler field theory which has an interesting relationship with the black hole interior action \eqref{new_lapse_action} is the 2-dimensional BMS--Liouville theory of \cite{Merbis:2019wgk,Barnich:2017jgw,Barnich:2012rz}. In terms of the scalar $\xi$ and the connection $\varphi$ its action is\footnote{The fields now depends on two variables, but the symmetries are still in the form \eqref{phixireparam}, with 
\be
X(u,\tau) = Y(\tau)\,,\q\q \alpha(u,\tau)=T(\tau) + u \, \f{\de Y(\tau)}{\de \tau}\,.   
\ee
Usually the $\tau$ coordinate is compactified and represented by an angle $\phi$, but here we keep the name $\tau$ for a better comparison with our case.}
\be\label{BMS Liouville}
\cS_\rm{BMS} = \iint \de u\, \de \tau \big[ \partial_\tau\xi\, \partial_u \varphi - (\partial_\tau \varphi)^2 \big]\,.
\ee
Performing the Hamiltonian analysis of this action with respect to the new variable $u$, we find that the conserved charges generating the BMS$_3$ symmetries are given by
\be\label{BMS Liouville charge}
\cQ= \int \de \tau \big[J X+P \alpha \big],
\ee
where $J$ and $P$ are precisely the quantities \eqref{new J P}, whose vanishing is equivalent to the equations of motion. This can also be seen by computing the charges as the contraction of the transformations \eqref{phixireparam} with the symplectic structure $\omega=\delta(2\varphi-\xi)\delta\dot{\varphi}$ coming from \eqref{BMS Liouville}. As usual, one interprets this expression for the charges as a pairing between elements $(X,\alpha)$ in the adjoint representation and elements $(J,P)$ in the coadjoint representation. This particular relationship between the charges of the 2d BMS$_3$ invariant theory and the equations of motion of the 1d model \eqref{new_lapse_action} becomes clearer if we look at the possibility of embedding our system in the coadjoint representation of BMS$_3$. This is the subject of the next section.

\section{Coadjoint representation of BMS$_3$ and central charges}
\label{coadj_section}

The understanding of coadjoint orbits of the symmetry groups is crucial in order to go towards the quantum theory. In particular the classification of the orbits of semi-direct product groups by their little groups will naturally leads to irreducible representation of the full group. In the case of the BMS$_3$ group, a detailed discussion is presented in \cite{Oblak:2016eij}. In appendix \ref{BMS_appendix} we report the key ingredients about the coadjoint orbits which we need in the present work.

The main goal of this section is to explain how the identification of the gravitational fields with the $\mathfrak{bms}_3$ Lie algebra elements naturally leads to the introduction of 2-forms on the circle that transform as covectors under the group action. Moreover, we show here that the action itself can be seen as a bilinear form between vectors and covectors. For this, let us first consider the equations of motion for the configuration fields $V_i$, written in the form
\bsub\label{coadj_vector}
\be
P&= \f{\ddot V_1}{V_1}-\f{\dot{V}_1^2}{2V_1^2}= \f{1}{2}\left (\dot \varphi^2 - 2\ddot \varphi\right )\,, \label{supermomenta} \\
J&= -\f{\dot{V}_2 \dot{V}_1 }{V_1^2} + \f{\ddot{V_2}}{V_1}+\f{V_2}{V_1}\left ( \f{\dot{V}_1^2}{V_1^2} - \f{\ddot V_1}{V_1}\right )=\ddot{\xi}-\dot \xi \dot \varphi\,.
\ee
\esub
A straightforward calculation reveals that $J$ and $P$ transform exactly as in the coadjoint representation of the centrally-extended group $\text{BMS}_3$, i.e. as \eqref{BMS_coadj_apppendix}, with central charges given by\footnote{The central charge $c_2$ appears here as a dimensionless quantity, while in 3-dimensional gravity it is usually normalized to $c_2 = 3/G$. Nevertheless, by a constant and dimensionfull rescaling of $P$ in \eqref{supermomenta} we can arbitrarily change the value of $c_2$ (and its dimension). The normalization in 3-dimensional gravity comes from the fact that the zero mode of the supermomentum represents the Bondi mass aspect, while here the on-shell value of $P$ is zero, so its dimension is not important. We keep the choice leading to ligher notations.}
\be
c_1= 0\q\q c_2 = 1.
\ee 
More precisely, defining the coadjoint action as
\be
\text{Ad}^*_{f^{-1},g} (J,P)\coloneqq \left (\text{D}_{f^{-1}}\circ \text{T}_{g}\right ) (J,P) \eqqcolon(\tilde J,\tilde P)\,,
\ee
we find the transformation laws
\be
\label{BMS_coadj}
\tilde P=\dot{f}^2(P \circ f) - \,\text{Sch}[f]\,,
 \q\q
 \tilde{J} =   \dot{f}^2\left(J+g\dot{P}+2\dot{g} P - g^{(3)}\right )\circ f\,.
\ee

As a remark, one can note that the other central charge can be switched on by the shift $J \to J+cP$, which leads to $c_1 =c$. This shift can be obtained by adding a term $c P V_1$ to the action \eqref{new_lapse_action}. Remarkably, this shift does not affect the equations of motion and only modifies the symplectic potential. This comes from the fact that
\be\label{delta PV1}
\delta(PV_1)=P\delta V_1+\f{\de}{\de\tau}\left(\delta\dot{V}_1-\f{\delta V_1\dot{V}_1}{V_1}\right),
\ee
so that the variation of the new Lagrangian is $\delta(\cL_0+cPV_1)=(J+cP)\delta V_1+P\delta V_2+\de_\tau\bar{\theta}$, where $\bar{\theta}$ is the sum of the boundary term \eqref{delta PV1} and the potential \eqref{V potential} coming from $\cL_0$. The equations of motion therefore combine to give once again $J=0$ and $P=0$. It is very intriguing that this mechanism is the same as in 3-dimensional gravity, where the BMS$_3$ central charge $c_1$ (or a chiral mismatch $c_+\neq c_-$ between the two Brown--Henneaux central charges in the AdS$_3$ case) can be switched on by adding a Chern--Simons and a torsion term (together these constitute Witten's exotic Lagrangian \cite{MR974271}) to the first order Lagrangian, without however modifying the equations of motion \cite{Blagojevic:2006hh,Geiller:2020edh,Geiller:2020okp}.

By adding an appropriate boundary term to the initial action \eqref{new_lapse_action}, it is also possible to write a new action as the bilinear form between vectors and covectors. This rewriting is done by defining
\be
\label{BMS_geometric_action}
\cS[V_i]&\coloneqq\cS_0[V_i] - \f{1}{G}\int\de \tau \f{\de}{\de \tau}\left (\f{\dot{V}_1V_2}{V_1} \right ) \cr
&=  \f{1}{G}\int \de \tau \left[\f{ V_2 \dot{V}_1^2}{2 V_1^2} -  \f{\dot{V}_1 \dot{V}_2}{V_1} + \ddot V_2 \right]\cr
&=  \f{1}{G}\int \de \tau \big[J V_1 + P V_2 \big].
\ee 
By construction this action leads to the same equations of motion as $\cS_0$. It is however remarkable that the explicit variation is again of the form
\be
\label{BMS_geometric_action_variation}
\delta\cS=\f{1}{G}\int \de \tau\,\big[ J\delta V_1+P\delta V_2+\de_\tau\tilde{\theta}\,\big]\,,
\ee
even though $J$ and $P$ (expressed in terms of the $V_i$'s as in \eqref{coadj_vector}) have of course been varied in \eqref{BMS_geometric_action} to obtain this expression! Consistently, the only difference with \eqref{variation_action_0} is the form of the symplectic potential (which we have not displayed here because it won't play a role in what follows).

The result is the 1-dimensional action \eqref{BMS_geometric_action} for the black hole interior in proper time gauge. This action, which is \textit{not} BMS$_3$ invariant, is however equivalent to the charge expression \eqref{BMS Liouville charge} for the 2-dimensional BMS$_3$ invariant theory \eqref{BMS Liouville}, provided we identify the infinitesimal parameters $(X,\alpha)$ in the charge with the algebra element $(V_1,V_2)$. We note that this new action transforms under BMS$_3$ as
\be
\label{transfo_geometric_action}
\text{Ad}^*_{(f^{-1},g)} \cS[V_i,\tau] = \cS[V_i,\tilde {\tau}] -\f{1}{G}\int \de \tilde \tau\left [(V_2-\dot{g}V_1 +g \dot{V}_1)\,\text{Sch}[f^{-1}] - V_1 g^{(3)} -\f{\de}{\de \tilde \tau} (g V_1 P) \right ],
\ee
which as expected differs from \eqref{transfo_action} only by a boundary term.

We can now give a nice interpretation of the Poincar\'e symmetry subgroup as an orbit stabilizer. So far we have embedded the fields $V_1$ and $V_2$ within the $\bms_3$ Lie algebra. We can build corresponding group elements by considering the functions
\be
f(\tau)\coloneqq  \int^\tau \f{\de s}{V_1(s)}\,,\q\q g(\tau) \coloneqq  \int^\tau \de s \,\left (\f{V_2}{V_1} \right )\circ f^{(-1)}(s)\,.
\ee
Transforming these quantities under BMS$_3$ reveals that they are indeed group elements. If we now insert these functions into the transformation law \eqref{BMS_coadj} for the covectors, starting from the point $(P=0,J=0)$ we find that the transformed pair $(\tilde J, \tilde P)$ is precisely equal to the right-hand side of \eqref{coadj_vector}. In other words, for the two functions $(f,g)$ given above we have
\be
-\text{Sch}[f]=\f{\ddot V_1}{V_1}-\f{\dot{V}_1^2}{2V_1^2}=P\,,\q\q -\dot{f}^2\, g^{(3)}\circ f = -\f{\dot{V}_2 \dot{V}_1 }{V_1^2} +\f{\ddot{V_2}}{V_1} +\f{V_2}{V_1}\left ( \f{\dot{V}_1^2}{V_1^2} - \f{\ddot V_1}{V_1}\right )=J.
\ee
This shows that demanding that the fields satisfy the equations of motion is equivalent to the condition that ($P=0,J=0$) be preserved under \eqref{BMS_coadj}, which means in turn that $(f,g)$ belongs to the stabilizer of the orbit. In general the coadjoint orbit of BMS$_3$ is defined by the Virasoro orbit spanned by the supermomentum $P$. These orbits are usually classified for the diffeomorphism group of the circle, while here we are dealing with functions on the real line since $\tau\in \R$. We can however compactify the proper time with \textit{e.g.} the reparametrization $\tau\to \arctan \theta/2$. In particular, this maps the point $(P=0,J=0)$ to $(P=-1/2,J=0)$, i.e. the constant representative of the so called \textit{vacuum} orbit \cite{Oblak:2016eij}. The stabilizer of this orbit is known to be ISO(2,1), which of course is precisely the symmetry group of our theory.

\section{Discussion}

In \cite{Geiller:2020xze} we have studied the ISO(2,1) (i.e. $(2+1)$-dimensional Poincar\'e) symmetry of the homogeneous action \eqref{new_lapse_action} describing the black hole interior spacetime, that is
\be
\cS_0
=\f{1}{G}\int \de \tau \, \f{ \dot{V}_1 ( V_2 \dot{V}_1 - 2  V_1  \dot{V}_2)}{2V_1^2 }\,,
\tag{\ref{new_lapse_action}}
\ee
where $V_1$ and $V_2$ are components of the 4-metric.
This previous work has hinted at the fact that the infinite-dimensional extension BMS$_3$ of the Poincaré group could also play a role in understanding the origin and the role of this symmetry. In the present paper, we have pushed along this direction and studied the action \eqref{new_lapse_action}
for the  geometry inside the black hole
from the point of view of BMS$_3$.

For this, we have shown in section \ref{sec:bms_adjoint} how finite and infinitesimal BMS$_3$ transformations act on \eqref{new_lapse_action}. These transformations do not leave the action invariant, and only the subgroup of transformations corresponding to Poincar\'e does. Nevertheless, in the case of the supertranslations, one can interpret the non-invariance of the action as creating a term corresponding to a cosmological constant. Acting with a further supertranslation then preserves the form of the Lagrangian while however changing the value of the cosmological ``constant''. We have shown that even if the BMS transformations are not strictly speaking symmetries of the action, they nevertheless admit integrable generators on the phase space of the theory. Their charge algebra, given by \eqref{poisson_bms}, does however fail to reproduce the centrally-extended $\mathfrak{bms}_3$. This is to be expected since these are indeed not symmetries of the theory. We have explained how a fully BMS-invariant action \eqref{invariant_1d_action} can be obtained by promoting the BMS$_3$ group elements to dynamical variables. This higher order action was then studied in appendix \ref{Full_invariant}. Aside from this 1-dimensional invariant action, there are 2-dimensional geometrical actions typically arising in studies of the boundary dynamics of 3-dimensional gravity \cite{Barnich:2012rz,Barnich:2013yka,Merbis:2019wgk}. This is for example the BMS$_3$ invariant action \eqref{BMS Liouville}. Intriguingly, the charges of this 2-dimensional action are written in terms of coadjoint vectors which turn out to be precisely the equations of motion of our starting point action \eqref{new_lapse_action}. This has motivated the study of the coadjoint representation of BMS$_3$ in section \ref{coadj_section}. Note that for the action \eqref{invariant_1d_action} the BMS symmetries are \textit{gauge symmetries}, while in the case of \eqref{BMS Liouville} they are \textit{physical symmetries}.

We have then shown that the action \eqref{new_lapse_action} for the black hole interior can be rewritten (up to a boundary term innocent for the equations of motion) as the compact geometric action
\begin{equation}
\cS=  \f{1}{G}\int \de \tau \left[\f{ V_2 \dot{V}_1^2}{2 V_1^2} -  \f{\dot{V}_1 \dot{V}_2}{V_1} + \ddot V_2 \right]
=  \f{1}{G}\int \de \tau \big[J V_1 + P V_2 \big]\,.
\tag{\ref{BMS_geometric_action}}
\end{equation}
This is made possible by the fact that $J$ and $P$ (which are defined in terms of $V_1,V_2$ and their time derivatives)
are coadjoint vectors under BMS$_3$ with central charge $c_2=1$, while the configuration variables $V_1$ and $V_2$ are elements of the Lie algebra $\mathfrak{bms}_3$.
Remarkably, the variation of this action \eqref{BMS_geometric_action} leads to the field equations $J=P=0$, which are exactly equivalent to the original equations of motion.
%
This property of the geometric action also enables to turn on the other central charge $c_1$ by adding a term to the Lagrangian without however modifying the equations of motion. In this construction, the Poincar\'e subgroup corresponds to the stabilizer of the vacuum orbit of BMS$_3$.

These results show that even if the action \eqref{new_lapse_action} is only invariant under Poincar\'e symmetries, there is a meaningful way in which one can understand this invariance as a broken BMS$_3$ symmetry. This confirms the intuition that this latter symmetry group does indeed plays a foundational role in the physics of the black hole interior, although  it does  not appear here in the more usual way as an asymptotic or horizon boundary symmetries. It is also intriguing, as shown in appendix \ref{other_slicing}, that other gravitational systems in 4, 3, and 2 spacetime dimensions involve the same action \eqref{new_lapse_action} and therefore admit the same symmetry structure.

In order to investigate further the origin and the physical role of these symmetries (both the Poincar\'e and the extended BMS one), there are several directions to be developed:
\begin{itemize}
\item One should investigate to which physical systems correspond the other BMS coadjoint orbits. A related question is how the phase space structure of these coadjoint orbits can be used to define a quantization of the system. This can already be investigated in terms of Poincar\'e representations \cite{francesco-rep}, but the embedding into BMS could allow to shift the Casimirs and to describe new physical processes related to the dynamics of the black hole interior.
\item An intriguing question remains that of the relationship between the BMS group appearing in the present context and that appearing for asymptotic or near-horizon symmetries. It is particularly interesting how the Poincar\'e symmetry of \eqref{new_lapse_action} can be seen as a broken BMS symmetry as in the case of SL$(2,\mathbb{R})$ and the Schwarzian action in JT gravity.
\item A related question is that of understanding BMS-invariant actions in mechanics and field theory. It could be that \eqref{new_lapse_action} descends from a higher-dimensional BMS-invariant action with a gauge-fixing and a dimensional reduction.
\item We have seen that the BMS transformations do not leave the action invariant, but that in a sense (which is clear for supertranslations) they generate new terms which can be interpreted e.g. as a cosmological constant. Then the BMS transformations generate a flow in a space of theories, and preserve the form of the action while changing its couplings. It would be very interesting to investigate the extension of Noether's theorem(s) to this type of transformations (which are very similar to renormalization flow transformations).
\item Finally, another possible generalization of this work, motivated in parts by the symmetry structure of 3-dimensional gravity, is to study an extension of \eqref{new_lapse_action} which admits as a symmetry group $\text{SL}(2,\mathbb{R})\times\text{SL}(2,\mathbb{R})$ instead of Poincar\'e's $\text{SL}(2,\mathbb{R})\ltimes\R^3$, and therefore involves a double Virasoro algebra instead of BMS$_3$. In the case of 3-dimensional gravity, this is done by turning on the cosmological constant, while in the case of the black hole interior, this is evidently not so. There are however other simple terms to be added to the action \eqref{new_lapse_action} in order to obtain this deformation to double Virasoro. The gravitational interpretation of these terms in the present context is however not yet clear, but is the subject of ongoing work.
\end{itemize}


\newpage
\appendix

\section{The Virasoro and BMS$_3$ groups}
\label{BMS_appendix}

In this appendix we recall basic facts about the Virasoro and the BMS$_3$ groups, as well as their centrally-extended version. We give in particular the form of the coadjoint action used in the main text.

\subsection{The Virasoro group}

We recall here some basic properties of the Virasoro group as the central extension of the group of diffeomorphisms of the circle $S^1$. We denote by $\text{Vect}(S^1)$ the space of vector fields on the circle endowed with the bracket
\be
\big[f(\theta)\,\partial_\theta ,\,g(\theta)\,\partial_\theta\big]=(f g' -g f')\partial_\theta \q\q 
f\,\partial_\theta ,\,g\,\partial_\theta \in \text{Vect}(S^1)\,,
\ee
with $f$ and $g$ periodic functions of $\theta$ with period $2 \pi$. A standard basis is given by the Fourier modes $\ell_n\coloneqq  i e^{in \theta}\partial_\theta$, with the commutation relations $[\ell_n,\ell_m] = -i(n-m) \ell_{n+m}$. This is the Witt algebra. Moreover, the vector space $\text{Vect}(S^1)$ can been seen as the space of generators of the (orientation preserving) diffeomorphisms on the circle, denoted by $\text{Diff}^+(S^1)$. Conversely, the diffeomorphisms can be endowed with a Lie group structure, whose algebra is $\text{Vect}(S^1)$.

\subsubsection{Tensor densities on $S^1$}

We denote by $\cF_\lambda$ the space of tensor densities of degree $\lambda\in \R$. It consists of elements of the form $\alpha=\alpha(\theta) \de \theta^\lambda$. Diffeomorphisms act by the adjoint action as
\be
\alpha(\theta) \xrightarrow{\varphi\, \in\, \text{Diff}^+(S^1)}  \text{Ad}^*_{\varphi^{-1}} \alpha(\theta) = \left (\varphi'(\theta)\right )^\lambda \alpha\left (\varphi(\theta)\right ).
\ee
The infinitesimal version gives the adjoint action of $\text{Vect}(S^1)$ on densities, i.e. the Lie derivative of the differential along a vector field. Taking $\varphi=\mathbb{I}+\epsilon X$, with $\epsilon \to 0$ and $X=X(\theta)\partial_\theta$, we get
\be
L^{(\lambda)}_X \alpha\coloneqq \text{ad}^*_{X} \alpha = X \alpha'+\lambda X' \alpha.
\ee
We can naturally define a bilinear form on $\cF_\lambda \times \cF_{1-\lambda}$ which is invariant under Lie derivative as
\be
\braket{\alpha,\beta}\coloneqq \int_{S^1} \alpha \otimes \beta\ \q\q  \forall\ \alpha \in \cF_\lambda\,,\ \beta \in \cF_{1-\lambda}\,.
\ee
We then identify the dual of $\text{Vect}(S^1)$ with the space of quadratic differentials $p= p (\theta)\de \theta^2 \in \cF_2$, and we have
\be
\braket{p,v}\coloneqq \int_{0}^{2\pi}\de \theta\, p(\theta) v(\theta) \q\q \forall\ v \in \cF_{-1}\,.
\ee
This means that the action of the vector fields on $\cF_2$ coincides with the co-adjoint action, i.e.
\be\label{coadjoint_def}
\braket{\text{ad}^*_{X}(p),Y}=-\braket{p,[X,Y]}\q  \forall\ X,Y \in \cF_{-1}\,,\ p \in \cF_{2}\,.
\ee

\subsubsection{Central extension}

We recall that the only non-trivial cocycle on $\text{Vect}(S^1)$ is given by the Gelfand--Fuchs 2-cocycle
\ba
\omega : \text{Vect}(S^1) \times \text{Vect}(S^1) &\to& \R\\
(X,Y) &\mapsto& \int \de \theta\, X' Y'' = \f 1 2 \int \de \theta (X' Y''-X''Y')\,.  \notag
\ea
The Virasoro algebra is the central extension of the algebra of vector fields, defined on the vector space $\widehat{\text{Vect}(S^1)}\coloneqq\text{Vect}(S^1) \oplus \R$, with the bracket
\be
\big[(X,a),(Y,b)\big]\coloneqq \big([X,Y], \omega(X,Y)\big) \q\q \forall\ a,b \in \R\,,\ X,Y \in \cF_{-1}\,.
\ee
The corresponding group is introduced via the Bott--Thurston 2-cocycle in $\rm {Diff}^+(S^1)$, which is
\ba
B : \text{Diff}^+(S^1) \times \text{Diff}^+(S^1) &\to& \R\\
(\varphi,\psi) &\mapsto&\f{1}{2} \int \de \theta \log (\varphi' \circ \psi) \left  (\log(\psi')\right )'\,. \notag
\ea
We then define the Virasoro group as $\widehat{\text{Diff}^+}(S^1)=\text{Diff}^+(S^1) \times \R$ equipped with the group law
\be
(\varphi,a) \circ (\psi,b) = \big(\varphi \circ \psi, b + B(\varphi,\psi)\big) \,.
\ee
Consistently, we recover the Virasoro algebra by the observation that the infinitesimal limit of $B$ is $\omega$. Indeed, defining the flows $\varphi_t$ and $\psi_s$ corresponding respectively to the vector fields $X(\theta)\partial_\theta$ and $Y(\theta)\partial_\theta$, we have
\be
\omega(X,Y) = \f{\de^2}{\de s\,\de t}  \big (B(\varphi_t,\psi_s)-B(\psi_s,\varphi_t)\big ) \Big|_{t=s=0}\,.
\ee

\subsubsection{Coadjoint action}

With the extended Lie bracket and the bilinear form at our disposal, we can compute the coadjoint action of the Virasoro algebra on quadratic forms. Taking the adjoint action of the algebra on itself, we define its coadjoint action as in \eqref{coadjoint_def}. A straightforward calculation leads to
\be
\text{ad}^*_{(X,a)} (p,c) = \big(p'X +2 X'p - c X^{(3)} ,0\big) \q\q \forall\ X \in \cF_{-1}\,,\ p \in \cF_{2}\,, \ a,c \in \R\,.
\ee 
Its exponential gives the group coadjoint action
\be
\text{Ad}^*_{(f^{-1},a)} (p,c) = \big(p \bullet f - c \,\cS[f],\,c \big) \q\q \forall\ f \in \rm {Diff}^+(S^1)\,,\ p \in \cF_{2}\,, \ a,c \in \R\,,
\ee
where
\be
p \bullet f = f'(\theta)^2(p \circ f)\, \de \theta^2\,, \q\q \cS[f]= \text{Sch}[f]\de \theta^2= \left [\f{f^{(3)}}{f'}-\f{3}{2}\left (\f{f''}{f'}\right )^2\right ]\, \de \theta^2\,.
\ee
Finally, introducing $h=f'$, we recall that the Schwarzian derivative satisfies
\bsub
\be
\text{Sch}[f]&=\f{h''}{h}-\f{3}{2}\left (\f{h'}{h}\right )^2 = (\log h)'' -\f{1}{2} \big((\log h)'\big)^2\,,\\
\text{Sch}[f\circ g]&=\text{Sch}[g] +(g')^2\,\text{Sch}[f]\circ g\,,\\
\text{Sch}[f^{-1}]&=-\f{\text{Sch}[f]}{h^2}\,.
\ee
\esub

\subsection{The BMS$_3$ group}

The centrally-extended BMS$_3$ group is the semidirect product, under the adjoint action, of the Virasoro group and its algebra seen as an Abelian vector group. This is
\be
\widehat{\text{BMS}_3}=\widehat{\text{Diff}^+}(S^1) \ltimes_\text{Ad} \widehat{\text{Vect}}(S^1)_\text{ab}\,.
\ee
We will (abusively) refer to the first factor as superrotations, and to the second one as supertranslations. Its elements are quadruples $(f,a;g,b)$ with $f \in \text{Diff}^+(S^1)$, $g \in \text{Vect}(S^1)_\text{ab}$, and $a,b \in \R$. The adjoint action of the Virasoro group reads
\be
\text{Ad}_f (g,b)=\left ( (g f') \circ f^{-1}, b - \int \de \theta\, \text{Sch}[f] g  \right )\,.
\ee
The corresponding algebra is
\be
\widehat{\mathfrak{bms}_3}=\widehat{\text{Vect}}(S^1) \oplus_\text{ad} \widehat{\text{Vect}}(S^1)_\text{ab}\,,
\ee
whose elements are again quadruples $(X,a; \alpha,b)$ with $X \in \text{Vect}(S^1)$, $\alpha \in \text{Vect}(S^1)_\text{ab}$, and $a,b \in \R$. The commutation relations are
\be
\label{BMS_algebra}
\big[ (X,a;\alpha,b), (Y,r;\beta,s) \big ]=
\big ([X,Y], \omega(X,Y); [X,\beta]-[Y,\alpha],\omega(X,\beta) -\omega(Y,\alpha) \big )\,.
\ee
As always, we can define an adjoint action of the group on its algebra by conjugation, which simply consists in exponentiating the commutation relation above. In the specific case studied here this gives
\be
\label{BMS_adjoint}
\text{Ad}_{(f; g)} (X;\alpha)=\left ( \text{Ad}_f X; \text{Ad}_f \alpha + [\text{Ad}_f X,g]\right )\,.
\ee
We are now interested in the coadjoint representation. The dual of the algebra $\widehat{\mathfrak{bms}_3}$ is the space $\widehat{\text{Vect}}(S^1)^* \oplus \widehat{\text{Vect}}(S^1)_\text{ab}^*$, paired with the algebra element via the bilinear form
\be
\lb (J,c_1;P,c_2),(X,a;\alpha,b) \rb\coloneqq \int_{0}^{2\pi} \left(JX+ P\alpha\right ) \de \theta +c_1 a +c_2 b\,.
\ee
This leads to the coadjoint representations of $\widehat{\text{BMS}_3}$, which is
\be
\text{Ad}^*_{f^{-1},g} (J,c_1;P,c_2) =\left (\tilde J,c_1; \tilde P,c_2 \right)
\ee
with
\be
\label{BMS_coadj_apppendix}
\tilde P=f'^2(P \circ f) - c_2 \,\text{Sch}[f]\,,\q\q \tilde{J} =   f'^2\left(J+gP' +2 g'P - c_2 g^{(3)}\right )\circ f - c_1 \,\text{Sch}[f]\,.
\ee
This is what we use in the main text, for example in \eqref{BMS_coadj}, where there however the prime $'$ has become dot $\cdot$ since the variable is the proper time $\tau$ instead of the angle $\theta$.

\section{Other gravitational systems}
\label{other_slicing}

In this section we briefly present some other gravitational systems which exhibit the same algebraic properties (encoded in the Poincar\'e symmetry and its embedding into BMS$_3$) as the action $\cS_0$ which was our starting point.

\subsection{4d black hole in Eddington--Finkelstein coordinates}

The first example is the 4-dimensional black hole spacetime written in Eddington--Finkelstein null coordinates. This is described by the line element
\be 
\de s^2_\text{EF} = \f{8V_2(r)}{V_1(r)} \de u^2 + 4 L_0\, \de u\,\de r  + V_1(r) \de \Omega^2\,.
\ee
Up to a boundary term, with $\tau$ replaced by $r$, and with the non-compact coordinate $u$ regularized to $u \in[0,L_0]$, the Einstein--Hilbert action evaluated on this line element gives $\cS_0$, i.e.
\be
\cS_\text{EH}^{(\text{EF})}= \f{1}{G}\int \de r \,\left [L_0^2 +  \f{ \dot{V}_1 ( V_2  \dot{V}_1 - 2  V_1 \dot{V}_2)}{2V_1^2 } + \f{\de}{\de r}\left(\f{V_2 \dot{V}_1 + V_1 \dot{V}_2}{V_1} \right )\right ]\,,
\ee 
where the dot denotes derivative with respect to $r$.

\subsection{3d BTZ black hole interior}

Going lower in dimensions, it is interesting that the 3-dimensional BTZ black hole, which is defined in AdS spacetime, also leads to the action $\cS_0$. To see this, we consider the 3-dimensional line element
\be\label{KS3d}
\de s^2_\text{BTZ}
&=-N^2\de t^2+\f{2 V_2}{\sqrt{V_1}}\de r^2+\sqrt{V_1}(\de\phi+N_\phi\de r)^2,\cr
&=-N^2\de t^2+\left(\f{2V_2}{\sqrt{V_1}}+N_\phi^2\sqrt{V_1}\right)\de r^2+2N_\phi\sqrt{V_1}\de r\,\de\phi+\sqrt{V_1}\de\phi^2,
\ee
with coefficients depending only on $t$, and where we have included angular momentum $N_\phi$. With this line element we find $\sqrt{-g}=N\sqrt{2 V_2}$ and
\be
\sqrt{-g}\,R=\f{V'_1(V_2V'_1-2V_1V'_2)+2V_1(V_1N_\phi')^2}{4\sqrt{2}NV_1^2\sqrt{V_2}}+\f{\de}{\de t}\left(\f{\sqrt{2}V'_2}{N\sqrt{V_2}}\right).
\ee
We remark that the 3-dimensional Lagrangian has only a kinetic term and no potential. This is feature of 3-dimensional gravity, where the potential term is actually provided by the cosmological constant volume term. Dropping the boundary term above, and integrating over $r\in[0,L_0]$ and $\phi\in[0,2\pi]$ we find
\be
\cS_\text{3d}=\f{1}{2\pi}\int_M\de^3x\,\sqrt{-g}\,(R+2\Lambda)=\int_\mathbb{R}\de t\,L_0\left[\f{V'_1(V_2V'_1-2V_1V'_2)+2V_1(V_1N_\phi')^2}{4\sqrt{2}NV_1^2\sqrt{V_2}}+\Lambda N\sqrt{8V_2}\right].
\ee
Gauge fixing the lapse to introduce proper time $\de \tau  = \f{N\sqrt{8V_2}}{L_0} \de t$, we find once again the action $\cS_0$ up to some differences: 
\begin{itemize}
\item The constant term $L_0^2$ is now multiplied by the cosmological constant $\Lambda$.
\item There is an additional kinetic term depending on the angular momentum. This term is actually covariant under the BMS transformations, provided that $N_\phi$ is a scalar under superrotations and does not transform under supertranslations, i.e. 
\be
\rm{Ad}_{f,g} N_\phi(\tau) = N_\phi \circ f^{-1}.
\ee
\end{itemize}

\subsection{2d dilatonic gravity}

Finally, another example leading to $\cS_0$ can be found in 2-dimensional dilatonic gravity. In this context, we consider the 2-parameter family of actions
\be
\cS_\text{2d}=\int \de x^2\, \sqrt{-g}\,\Big(\Phi^2R+\alpha\Phi^{\alpha-2}+2\beta(\nabla\Phi)^2\Big).
\ee
This includes the spherically-symmetric section of 4d gravity $(\alpha=2,\beta=1)$, the CGHS model $(\alpha=4,\beta=2)$, and JT gravity $(\alpha=4,\beta=0)$. Let us then consider a gauge where the dilaton depends only  on $r$, and with 2-dimensional metrics of the form
\be
\de s^2= A(r) \de u^2 +2 B(r) \de u\, \de r\,.
\ee
If we now choose the fields as
\bsub
\be
A(r) &= \f{\alpha-\beta}{\alpha} V_1^{\f{\alpha}{2(\alpha-\beta)}} V_2^{-\f{\beta}{\alpha-\beta}}\,,\\
B(r) &= \f{L_0}{\alpha} V_1^{\f{\alpha-2}{2(\alpha-\beta)}} V_2^{\f{2-\alpha}{\alpha-\beta}}\,,\\
\Phi(r) &= V_1^{-\f{1}{2 \alpha - 2 \beta}} V_2^{\f{1}{\alpha - \beta}}\,,
\ee
\esub
the dilatonic action reduces again to $\cS_0$ up to a boundary term if we choose $u \in[0,L_0]$, and replace $\tau$ with $r$. More precisely, we find
\be
\cS_\text{2d} = \int \de r \,\left [L_0^2 +  \f{ \dot{V}_1 ( V_2  \dot{V}_1 - 2  V_1 \dot{V}_2)}{2V_1^2 } + \f{\de}{\de r}\left(\alpha \f{V_2 \dot{V}_1}{2 V_1}- \beta \dot{V}_2 \right )\right ]\,,
\ee
where dot now denotes derivation with respect to $r$.

\section{Cosmological constant and scalar field}
\label{lambda_and_phi}

In this appendix we briefly explain how the action \eqref{new_lapse_action} is modified in the presence of a cosmological constant and an homogeneous scalar field. This shows that the contribution of these matter sources is proportional to $V_1$ and $V_2$ in the action, just like the extra terms which have been produced by the action of a finite BMS transformation in \eqref{transfo_action}.

Using the metric with redefined lapse $N_\text{p}$ as in \eqref{lapse Np}, working with proper time $\tau$ and introducing the IR cutoff $L_0$ to regulate the integration over $x$, the inclusion of a cosmological constant is done by adding to the Einstein--Hilbert action the volume term
\be
\cS_\Lambda = \f{1}{8 \pi G}\int \de^4 x\,  \sqrt{-g}\, \Lambda=\int \de \tau \, L_0V_1\Lambda\,.
\ee
Similarly, the contribution of a minimally-coupled scalar field $\phi$ in the spherically-symmetric and homogeneous spacetime \eqref{lapse Np} is described by the action
\be\label{scalar field in S}
\cS_\phi = -\f{1}{16 \pi G}\int \de^4x\, \sqrt{-g} \left ( \f{1}{2} (\nabla \phi)^2- \cV(\phi) \right )= \f{1}{2G}\int \de \tau \left (\dot \phi^2V_2  +L_0V_1 \cV(\phi)\right )\,.
\ee
Comparing these two actions with the transformation \eqref{transfo_action}, one can see that for pure supertranslations (i.e. with $f=\mathbb{I}$) we can interpret the third derivative of $g$ as an effective cosmological constant, while for pure superrotations (i.e. with $g=0$) we can interpret the term created by the superrotation as the kinetic term of a scalar field (up to a redefinition $\text{Sch}[f^{-1}]\mapsto\dot{\phi}^2$).

\section{1d BMS$_3$ invariant action}
\label{Full_invariant}

This appendix is devoted to the study of some properties of the full BMS invariant theory \eqref{invariant_1d_action}, which is
\be
\cS_\text{inv}[V_i,\Phi,\Psi]=\f{1}{G}\int  \de \tau\left [\f{ \dot{V}_1 ( V_2 \dot{V}_1 - 2  V_1  \dot{V}_2)}{2V_1^2 }+(V_2+ \dot \Psi V_1- \Psi \dot{V}_1) \text{Sch}[\Phi]  + V_1\Psi^{(3)}  \right ].
\ee
We first study the Lagrangian equations of motion and symmetries, and then perform the Hamiltonian analysis.

\subsection{Lagrangian analysis}

First of all we need to remark that the field $\Phi$ does not appear in the Lagrangian without derivatives, implying the existence of a ghost zero mode for $\Phi$. On the other hand, both $\Psi$ and $\Phi$ appear with third derivatives. Moreover, the action exhibits an infinite-dimensional gauge symmetry (under BMS$_3$), which will be reflected in the freedom to arbitrarily choose two of the fields in the solutions.

In spite of these peculiarities, the equations of motion takes nevertheless a remarkably simple form in terms of the coadjoint vectors defined in \eqref{J def} and \eqref{P def}. They are given by
\bsub\be
0\equiv \cE_{V_1} &= J + \Psi\,  \f{\de\,\text{Sch}[\Phi]}{\de \tau} + 2\dot \Psi\, \text{Sch}[\Phi] + \Psi^{(3)}\,,\label{eom_v1}\\
0\equiv \cE_{V_2} &= P+\text{Sch}[\Phi]\,.\label{eom_v2}
\ee\esub
The other two equations of motion arise as a consequence of these first two. Indeed, one an check that we have
\bsub\be
0\equiv\cE_{\Psi} &= - \f{1}{V_1} \f{\de}{\de \tau} \left( \cE_{V_2} V_1^2\right )  \,,\\
0\equiv \dot \Phi\cE_{\Phi} &= - \f{1}{V_1} \f{\de}{\de \tau} \left( \cE_{V_1} V_1^2\right ) -\f{1}{V_2} \f{\de}{\de \tau} \left( \cE_{V_2} V_2^2\right )+ \f{1}{\Psi} \f{\de}{\de \tau} \left[ \f{\Psi^2}{V_1} \f{\de}{\de \tau} \left( \cE_{V_2} V_1^2\right )\right ]\,.
\ee\esub
This makes manifest that the theory posses gauge symmetries, provided by the BMS$_3$ group. The equations of motion are solved by the 2-parameter family of solutions
\be
V_1 = \f{1}{\dot \Phi}\,,\q\q
V_2=\f{\mathfrak{c_1}+\mathfrak{c_2}\Phi}{\dot \Phi} - \f{1}{\dot\Phi}\f{\de}{\de \tau}\big(\Psi \dot \Phi\big) \,,
\ee
for some constant $\mathfrak{c}_i$'s. Moreover, we have that different solutions can be mapped into each other according to the monodromy condition satisfied by the supermomentum $P$ (one can compare the equations of motion with \eqref{BMS_coadj} in order to see this), meaning that choosing a particular solution is equivalent to picking up a point on a coadjoint orbit of BMS$_3$ entirely determined by the constants of integration $\mathfrak{c}_1$ and $\mathfrak{c}_2$. 

We can also compute the Noether charges associated to the BMS$_3$ symmetry. They are given by
\bsub
\be
\cD_X &= -X\left (V_1 \big(J - \Psi\, \dot P  + 2\dot \Psi \, \text{Sch}[\Phi]+ \Psi^{(3)}\big) +(V_2-2 \Psi \dot V_1) \big(P+\text{Sch}[\Phi]\big)\right )\,, \\
\cT_\alpha &= \alpha\, V_1 \big(P+\text{Sch}[\Phi]\big)\,,
\ee
\esub
that trivially vanishes on-shell, as expected for a gauge symmetry charges.

\subsection{Hamiltonian analysis}

Since the Lagrangian in \eqref{invariant_1d_action} is higher-order, in order to perform the Hamiltonian analysis we need to follow the Ostrogradski algorithm and introduce auxiliary fields with appropriate constraints. First, we therefore define
\be
\Phi_1 = \dot \Phi\,, \q\q \Phi_2 = \ddot \Phi\,,\q\q
\Psi_1 = \dot \Psi\,, \q\q \Psi_2 = \ddot \Psi\,.
\ee
With this, the action \eqref{invariant_1d_action} can then be rewritten with at most first order derivatives as
\be
\cS_\text{inv}&=\cS_0 + \f{1}{G} \int \de \tau \left[(V_2 + \Psi_1 V_1 - \Psi \dot{V}_1)\left(\f{\dot{\Phi}_2}{\Phi_1}-\f{3}{2}\left(\f{\Phi_2}{\Phi_1}\right)^2\right)+ V_1 \dot \Psi_2 \right.\cr
&\phantom{=\cS_0 + \f{1}{G} \int \de \tau \left[\right.}+\Pi(\dot \Phi - \Phi_1)+\Pi_1(\dot \Phi_1 - \Phi_2)+\Xi(\dot \Psi - \Psi_1)+\Xi_1(\dot \Psi_1 - \Psi_2) \Bigg]\,.
\ee
Here the second line is the set of constraints, imposed with Lagrange multiplier $(\Pi,\Pi_1,\Xi,\Xi_1)$, ensuring that the auxiliary fields can be eliminated in order to go back to the initial expression \eqref{invariant_1d_action}. As expected, because this theory has gauge symmetries it is a constrained system. In order to see this in the Hamiltonian formulation, we perform the Legendre transform and introduce a 16-dimensional phase space with canonical pairs
\be
\big\{(\Phi,\Pi),(\Phi_1,\Pi_1),(\Phi_2,\Pi_2),(\Psi,\Xi),(\Psi_1,\Xi_1),(\Psi_2,\Xi_2),(V_1,P_1),(V_2,P_2)\big\}\,,
\ee
with
\be
\lb\Phi_i,\Pi_i\rb=\lb\Psi_i,\Xi_i\rb=\lb V_i,P_i\rb=1\,.
\ee
The extended Hamiltonian is then found to be
\be
H&= H_0+ \dot \Phi_2 \chi_1 + \dot \Psi_2 \chi_2\,,\\
H_0&= -P_2 \left(V_1P_1+\f{1}{2} V_2 P_2 \right)+(V_2 + \Psi_1V_1+\Psi V_1P_2)\f{3}{2}\left(\f{\Phi_2}{\Phi_1}\right)^2+ \Pi \Phi_1+ \Pi_1 \Phi_2  + \Xi \Psi_1+ \Xi_1 \Psi_2\,,\notag
\ee
where $\dot \Phi_2$ and $\dot \Psi_2$ are Lagrange multipliers imposing the primary constraints
\bsub
\be
\chi_1&= \Pi_2 -\f{1}{\Phi_1}(V_2 + \Psi_1V_1+\Psi V_1P_2)\,,\\
\chi_2&= V_1 - \Xi_2\,.
\ee
\esub
We can now apply Dirac's algorithm and evolve the primary constraints with respect to the extended Hamiltonian. This evolution generates two secondary constraints given by
\bsub
\be
\chi_3&=\lb\chi_1,H_\text{tot}\rb=\f{1}{\Phi_1}\left(V_1P_1+V_2P_2-V_1\Psi_2+\f{1}{2}V_1P_2^2\Psi\right)-\f{2\Phi_2}{\Phi_1^2}(V_2+\Psi_1V_1+\Psi V_1P_2)-\Pi_1\,,\\
\chi_4&= \{\chi_2,H_\text{tot} \}= -P_2 V_1 +\Xi_1\,.
\ee
\esub
The evolution of these secondary constraints then leads to two tertiary constraints given by
\bsub
\be
\chi_5= \{\chi_3,H_\text{tot} \}&= \Pi-\f{1}{2\Phi_1}P_2\big(P_2 V_2-V_1 (P_2 \Psi_1-2 P_1 + 2 \Psi_2)\big)\cr
&\phantom{=\ } +\f{\Phi_2}{2\Phi_1^2}\big(2 P_2 V_2+  V_1 (2 P_1 + P_2^2 \Psi - 2 \Psi_2)\big)- \f{\Phi_2^2}{2\Phi_1^3}\big(V_2 + V_1 (P_2 \Psi + \Psi_1)\big)\,, \\
\chi_6= \{\chi_4,H_\text{tot} \}&=\f{1}{2}P_2^2 V_1 -\Xi\,, \\
H_\text{tot} &= H_0 + \mu_i \chi_i\,.
\ee
\esub
The algorithm then stops without giving any conditions on the Lagrange multipliers or on the fields. All the constraints are first class: their brackets weakly vanish, and for $\chi_5$ and $\chi_6$ they actually strongly vanish.

Now that we know the first class constraints, we need to find which combination generates the infinitesimal BMS$_3$ gauge transformations. A straightforward calculation shows that these Hamiltonian generators are given by
\bsub
\be
\cD_X&= X (-\Phi_1 \chi_5 + \chi_2 \dot \Psi_2 - \dot \Phi_2 \chi_1 + \Phi_2 \chi_3 + \Psi_1 \chi_6 - \Psi_2 \chi_4)+\\
\notag &\phantom{=\ }+ 
 \dot X (\Phi_1 \chi_3 - 2 \Phi_2 \chi_1 + \Psi_2 \chi_2 - \Psi \chi_6)+ \ddot X (\Psi \Phi_4 - \Psi_1 \chi_2 - \Phi_1 \chi_1)   - \Psi \chi_2 X^{(3)}\,,\\
\cT_\alpha&= -\alpha \chi_6 + \dot \alpha \chi_4 - \ddot \alpha \chi_2\,.
\ee
\esub
Finally, one can note that this is consistent with the Hamiltonian analysis of the initial action $\cS_0$, and that we obtain the $\text{ISO}(2,1)$ generators of the symmetry-broken theory in the limit
\be
(\Psi,\Pi_i,\Xi_i)\to 0\,,\q\q \Phi(\tau)\to \tau\,. 
\ee

\bibliographystyle{Biblio}
\bibliography{Biblio.bib}

\end{document}